\documentclass[english]{article}
\usepackage[utf8]{inputenc}
\usepackage{geometry}
\usepackage{babel}
\usepackage{amsthm}
\usepackage{amssymb}
\usepackage{graphicx}
\usepackage[unicode=true,pdfusetitle,
 bookmarks=true,bookmarksnumbered=false,bookmarksopen=false,
 breaklinks=false,pdfborder={0 0 1},backref=false,colorlinks=false]
 {hyperref}

\makeatletter
\theoremstyle{plain}
\newtheorem{thm}{\protect\theoremname}
\theoremstyle{plain}
\newtheorem{lem}[thm]{\protect\lemmaname}
\ifx\proof\undefined
\newenvironment{proof}[1][\protect\proofname]{\par
\normalfont\topsep6\p@\@plus6\p@\relax
\trivlist
\itemindent\parindent
\item[\hskip\labelsep
\scshape
#1]\ignorespaces
}{%
\endtrivlist\@endpefalse
}
\providecommand{\proofname}{Proof}
\fi

\usepackage{pstricks}

\makeatother

\providecommand{\lemmaname}{\inputencoding{latin9}Lemma}
\providecommand{\theoremname}{\inputencoding{latin9}Theorem}

\begin{document}
\global\long\def\R#1{R_{\left|#1\right\rangle }}

\global\long\def\bra#1{\left\langle #1\right|}

\global\long\def\ket#1{\left|#1\right\rangle }

\global\long\def\op#1#2{\left|#1\right\rangle \left\langle #2\right|}

\global\long\def\ip#1#2{\left\langle #1,#2\right\rangle }

\global\long\def\s#1{\sum_{#1\in\left\{  0,1\right\}  ^{k}}}

\global\long\def\I{\mathbf{I}}

\title{Asymptotically optimal approximation of single qubit unitaries by Clifford and T circuits using a constant number of ancillary qubits}

\author{{Vadym Kliuchnikov$^{1}$,  Dmitri Maslov$^{2,3}$, and Michele Mosca$^{4,5}$}\\
{\small\it $^1$ Institute for Quantum Computing, and David R. Cheriton School of
Computer Science} \\
{\small\it University of Waterloo, Waterloo, Ontario, Canada} \\
{\small\it $^2$ National Science Foundation} \\
{\small\it Arlington, Virginia, USA} \\
{\small\it $^3$  Institute for Quantum Computing, and Dept. of Physics \&
Astronomy} \\
{\small\it University of Waterloo, Waterloo, Ontario, Canada} \\
{\small\it $^4$  Institute for Quantum Computing, and Dept. of Combinatorics \&
Optimization} \\
{\small\it University of Waterloo,  Waterloo, Ontario, Canada} \\
{\small\it $^5$ Perimeter Insitute for Theoretical Physics} \\
{\small\it Waterloo, Ontario, Canada} \\}
\maketitle

\begin{abstract}
We present an algorithm for building a circuit that approximates single qubit unitaries with precision $\varepsilon$ using $O(\log(1/\varepsilon))$ Clifford and T gates and employing up to two ancillary qubits. The algorithm for computing our approximating circuit requires an average of $O(\log^{2}(1/\varepsilon)\log\log(1/\varepsilon))$ operations. We prove that the number of gates in our circuit saturates the lower bound on the number of gates required in the scenario when a constant number of ancillae are supplied, and as such, our circuits are asymptotically optimal. This results in significant improvement over the current state of the art for finding an approximation of a unitary, including the Solovay-Kitaev algorithm that requires $O(\log^{3+\delta}(1/\varepsilon))$ gates and does not use ancillae and the phase kickback approach that requires $O(\log^{2}(1/\varepsilon)\log\log(1/\varepsilon))$ gates, but uses $O(\log^{2}(1/\varepsilon))$ ancillae.   
\end{abstract}

\maketitle

\section{Introduction}

The efficient approximation of a unitary using a discrete universal gate set is crucial for building a scalable quantum computing device. Barenco {\em et al.} \cite{Barenco1995} showed that any unitary may be implemented by a circuit with CNOT and single qubit gates, effectively reducing the problem to that of the single qubit unitary synthesis/approximation.  A constructive answer to the question of how to approximate a single qubit unitary by a quantum circuit is given by the Solovay-Kitaev algorithm~\cite{kbook,skalg}.  While the Solovay-Kitaev algorithm may be applied to approximating multiple qubit/qudit unitaries by quantum circuits, in practice it remains most useful for single qubit approximations.

Technically, the problem of single qubit circuit synthesis is formulated as follows: given a discrete universal gate set or ``library'', find a sequence of gates in it that approximates a given unitary with precision $\varepsilon$.  Parameter $\varepsilon$ determines complexity of the resulting approximation.  

Computing an approximation using the standard version of the Solovay-Kitaev algorithm~\cite{skalg} takes $O(\log^{2.71}(1/\varepsilon))$ steps on a classical computer and the number of gates in the resulting quantum circuit is $O(\log^{3.97}(1/\varepsilon))$.  The best known upper bound on the circuit size resulting from the application of the Solovay-Kitaev algorithm is $O(\log^{3+\delta}(1/\varepsilon)),$ where $\delta$ can be chosen arbitrary small~\cite{kbook}.  From the other side, Harrow {\em et al.}~\cite{eda} show an $\Omega(\log(1/\varepsilon))$ lower bound on the number of gates in the approximating circuit.  A certain library of quantum gates that allows approximating a single qubit unitary to precision $\varepsilon$ with a circuit containing at most $O(\log(1/\varepsilon))$ gates is also reported in~\cite{eda}.  However, no efficient algorithm to construct a circuit meeting the lower bound in the number of gates is known.  Furthermore, the gate set used, $\frac{I+2i\{X,Y,Z\}}{\sqrt{5}}$, is not 
considered to be well-suited for a fault-tolerant implementation, in contrast to the Clifford and T library.  To the best of our knowledge, constructive saturation of the logarithmic lower bound in the Clifford and T library has not been shown yet, however, numerical evidence supports the theory that this is indeed the case~\cite{sqlg} (based on an exponential-time breadth first search algorithm).  Our result comes close to exactly meeting the lower bound---our gate count is logarithmic, $O(\log(1/\varepsilon))$, however, we use an additional resource in the form of at most two qubits initialized to the state $\ket{0}$. 

Allowing additional resources helps to achieve interesting improvements over the Solovay-Kitaev algorithm. For example, using a special resource state $\ket{\gamma}$ on $O(\log(1/\varepsilon))$ qubits allows to achieve the desired accuracy of approximation by a depth $O(\log(\log(1/\varepsilon)))$ circuit containing $O(\log(1/\varepsilon)$ gates~\cite{kbook}, also known as phase kickback algorithm.  However, the resource state preparation requires $O(\log^{2}(1/\varepsilon))$ ancillae qubits and a circuit of depth $O(\log^2(\log(1/\varepsilon)))$ containing $O(\log^{2}(1/\varepsilon)\log\log(1/\varepsilon))$ gates. Furthermore, exact preparation of the resource state $\ket{\gamma}$ is not possible using gates from the Clifford and T library and qubits initialized to $\ket{0}$~\cite{Kliuchnikov2012,mqes}. In comparison, in our work, we employ only two ancillae prepared in the simple state $\ket{0}$, and this results in achieving the approximating accuracy of $\varepsilon$ using a
circuit with $O(\log(1/\varepsilon))$ gates. Later in the paper we show the lower bound of $\Omega(\log(1/\varepsilon))$ on the number of gates required to approximate a unitary to the accuracy $\varepsilon$ using a fixed number of ancillae initialized to $\ket 0$ and any universal gate set. One other recent approach uses resource states~\cite{chem} and probabilistic circuits with classical feedback. The circuit itself, excluding state preparation, requires on average a constant number of operations and a constant number of ancilla qubits. The method requires precomputed ancillae in the states $R_Z(2^n\phi)H\ket{0}$ to implement $R_Z(2^m\phi)$. Our algorithm does not rely on the measurements and classical feedback, and our circuit is deterministic.  More importantly, our algorithm does not employ sophisticated ancilla states that, in turn, may require approximation, as they may not be possible to prepare exactly in the Clifford and T library~\cite{Kliuchnikov2012,mqes}.

In our previous work~\cite{Kliuchnikov2012}, we showed that any single qubit unitary with entries $u_{ij}$ in the ring $\mathbb{Z}\left[i,\frac{1}{\sqrt{2}}\right]$ can be synthesized exactly using single qubit Clifford and T gates.  Furthermore, we presented an asymptotically optimal algorithm for finding a circuit with the minimal number of Hadamard gates and asymptotically minimal total number of gates.  More precisely, if the square of the norm of an element of the single qubit unitary matrix, $|u_{ij}|^{2}$, can be represented as $(a+\sqrt{2}b)/2^{n}$, where $a$ and $b$ are integers such that $GCD(a,b)$ is odd, the total number of gates required to synthesize the unitary is in $\Theta(n)$.  This work opened the door for bypassing the Solovay-Kitaev algorithm for fast circuit approximation of single qubit unitaries by efficiently approximating arbitrary unitaries with unitaries over the ring $\mathbb{Z}\left[i,\frac{1}{\sqrt{2}}\right]$.  However, to date, no efficient ring round-off procedure was 
reported, and it remains an important open problem.

Giles and Selinger~\cite{mqes} recently found an elegant way to prove the conjecture formulated in~\cite{Kliuchnikov2012} stating that multiple qubit unitaries over the ring $\mathbb{Z}\left[i,\frac{1}{\sqrt{2}}\right]$ may be synthesized exactly using Clifford and T library.  In this paper, we employ some of their results to show that, by adding at most two ancilla qubits, we can achieve asymptotically optimal approximation of the single qubit unitaries in the Clifford and T library.

The significance of the improvement provided by our approach is best seen when, for a fixed precision $\varepsilon$, all of the approximating circuit parameters such as depth, the number of gates, and ancillae are added into one aggregate figure, such as, e.g., the product of the three of these parameters. 

\section{Main result}

We focus on the approximation of the following operator: 
\[
\Lambda(e^{i\phi}):\alpha\ket 0+\beta\ket 1\mapsto\alpha\ket 0+\beta e^{i\phi}\ket 1.
\]
We note that any single qubit unitary can be decomposed in terms of a constant number of Hadamard gates and $\Lambda(e^{i\phi})$ (see solution to Problem 8.1 in~\cite{kbook}).  Therefore, the ability to approximate $\Lambda(e^{i\phi})$ implies the ability to approximate any single qubit unitary.

There are two main steps in our algorithm: 
\begin{enumerate}
\item Find a circuit $C$ consisting of Clifford and T gates such that the
result of applying $C$ to $\ket{00}$ is close to $e^{i\phi}\ket{00}$. 
\item Apply circuit $C$ controlled on the first qubit to perform a transformation
close to: 
\[
\alpha\ket{000}+\beta\ket{100}\mapsto\alpha\ket{000}+\beta e^{i\phi}\ket{100}.
\]
\end{enumerate}

It can be observed that the net effect of such transformation may be described as the application of $\Lambda(e^{i\phi})$ to the first qubit. To accomplish the first step we approximate $e^{i\phi}\ket{00}$ with
a four dimensional vector $\ket v$ with entries in the ring $\mathbb{Z}\left[i,\frac{1}{\sqrt{2}}\right]$.
We then employ an algorithm for multiple qubit exact synthesis to find
a circuit $C$ that prepares $\ket v$ starting from $\ket{00}$ using at most one ancilla qubit. It was shown in~\cite{mtm} that any circuit
using Clifford and T gates can be transformed into its exact (meaning no further approximation is required) controlled
version with only a linear overhead in the number of gates, and using at most
one ancilla qubit in the state $\ket{0}$ that is returned unchanged.  Our analysis shows that, however, on this step we do not
need to use this additional ancilla.  The resulting total number of ancillae is thus at most two.

\subsection{Approximating $e^{i\phi}\ket{00}$}

The key is the reduction of the approximation problem to expressing
an integer number as a sum of four squares. In particular, we are looking for an approximation
of: 
\[
e^{i\phi}\ket{00}=\left(\cos\left(\phi\right)+i\sin\left(\phi\right)\!, 0, 0, 0\right)
\]
by a unit vector: 
\begin{eqnarray}
\ket v:=\frac{1}{2^{k}}\left(\left\lfloor 2^{k}\cos\left(\phi\right)\right\rfloor +i\left\lfloor 2^{k}\sin\left(\phi\right)\right\rfloor\!, 0, a+ib, c+id\right),\nonumber
\end{eqnarray}
\noindent where $k\in\mathbb{N}; a, b, c, d\in\mathbb{Z}$. Without loss of generality we can assume that $0\le\phi\le\frac{\pi}{4}$.
The power $k$ of the denominator determines precision of our approximation
and complexity of the resulting circuit. As $\ket v$ must be a unit
vector, the remaining four parameters ($a,b,c$, and $d$) should satisfy the integer equation: 
\[
a^{2}+b^{2}+c^{2}+d^{2}=4^{k}-\left\lfloor 2^{k}\cos\left(\phi\right)\right\rfloor^{2}-\left\lfloor 2^{k}\sin\left(\phi\right)\right\rfloor^{2}.
\]

Lagrange's four square theorem states that this equation always has
a solution. Furthermore, there exists an efficient probabilistic algorithm
for finding a solution. For the right hand side $M$ it requires on
average $O(\log^{2}(M)\log\log M)$ operations with integers smaller
than $M$.  It is described in Theorem 2.2 in~\cite{Rabin1986}. 
We get a reduction to such a simple Diophantine equation at the expense
of using two qubits instead of one. 

Furthermore, in estimating the classical complexity of the algorithm for finding the approximating circuit, we will rely on an observation that 
\[
4^{k}-\left\lfloor 2^{k}\cos\left(\phi\right)\right\rfloor^{2}-\left\lfloor 2^{k}\sin\left(\phi\right)\right\rfloor^{2} \leq 4\times 2^k + Const \in O(2^k).
\]

\subsection{Precision and complexity analysis}

Let us introduce $\gamma=\left(\left\lfloor 2^{k}\cos\left(\phi\right)\right\rfloor +i\left\lfloor 2^{k}\sin\left(\phi\right)\right\rfloor \right)/2^{k}$
and express $\ket v$ as:
\[
\ket v=\gamma\ket{00}+\ket 1\otimes\ket g.
\]
The application of the circuit $C$ controlled on the first qubit
will transform $\left(\alpha\ket 0+\beta\ket 1\right)\otimes\ket{00}$
into: 
\[
\alpha\ket{000}+\beta\gamma\ket{100}+\beta\ket{01}\otimes\ket g.
\]
The distance of the result to the desired state $\alpha\ket{000}+\beta e^{i\phi}\ket{100}$
is: 
\[
\sqrt{\left|\beta\left(e^{i\phi}-\gamma\right)\right|^{2}+\left|\beta\right|^{2}\left\Vert \ket g\right\Vert ^{2}}.
\]
By the choice of $\gamma$ we have $\left|\gamma-e^{i\phi}\right|\le\frac{\sqrt{2}}{2^k}$,
therefore the first term in the sum above is in $O(1/2^{2k})$. The
norm squared of $\ket g$ equals $1-\left|\gamma\right|^{2}$.
The complex number $\gamma$ approximates $e^{i\phi}$, and the distance
of its absolute value to identity can be estimated using the triangle
inequality: 
\[
\left|\left|\gamma\right|-\left|e^{i\phi}\right|\right|\le\left|\gamma-e^{i\phi}\right|.
\]
Therefore, $1-\left|\gamma\right|^{2}$ is in $O(1/2^{k})$. In summary,
the distance to the approximation is in $O(1/2^{0.5k})$. 

The same estimate is true if we consider the circuit $C$ as a part
of a larger system. In this case we should start with the state $\left(\alpha\ket{\phi_{0}}\otimes\ket 0+\beta\ket{\phi_{1}}\otimes\ket 1\right)\otimes\ket{00}$.
Similar analysis shows that the distance to approximation remains
$O(1/2^{0.5k})$.

As shown in~\cite{mqes}, it is possible to find a circuit that
prepares $\ket v$ using $O(k)$ Clifford and T gates (\cite{mqes}, Lemma 20 (Column lemma)). The classical complexity of constructing a quantum circuit implementing $\ket{v}$ is in $O(k)$.  In the controlled
version of this circuit the number of gates remains $O(k)$ (\cite{mtm}, Theorem 1). In summary, we need $O(\log(1/\varepsilon))$
gates to achieve precision $\varepsilon$.  The complexity of the classical
algorithm for constructing the entire approximating circuit is thus dominated by complexity of finding a solution to the Diophantine equation, which is in $O(\log^{2}(1/\varepsilon)\log\log(1/\varepsilon))$.

\subsection{How many ancillae are needed?}

A straightforward calculation shows that the number of ancillae used is three.  However, we can get around using only two ancillae.  To understand how, we need to go into the details of the proof of Lemma 20 (Column lemma) from~\cite{mqes}. It shows how to find a sequence of two-level unitaries of type $iX$, $T^{-m}(iH)T^m$, and $W$~\cite{mqes} and length $O(k)$ that allows to prepare a state with the denominator $2^{k}$. A controlled version
of the two level unitary is again a two level unitary. In~\cite{mqes}, Lemma 24, it was also shown that any such unitary required can be implemented using no extra ancillae. Therefore, the controlled version of the circuit $C$ will not use any additional ancilla and we need only two of them in total. 

\section{\label{sec:Lower-bound}Lower bound on the number of gates when ancillae are allowed}
\label{bound}
\begin{lem}
Let $G$ be a universal library, and
let $M_{V}$ be a set of unitaries, that simulate a unitary $V$ acting
on $n$ qubits, using $m$ ancillary qubits:
\[
M_{V}=\left\{ U\in\mathbb{U}\left(2^{m+n}\right)|U\left(\ket{0}\otimes\ket{\phi}\right)=\ket 0\otimes\left(V\ket{\phi}\right)\right\}.
\]
Then, for any $\varepsilon$ there always exists a unitary $V\left(\varepsilon\right)$
such that the number of gates from $G$ needed to construct a unitary
within the distance $\varepsilon$ to $M_{V\left(\varepsilon\right)}$
is in $\Omega(\log(1/\varepsilon))$. 
\end{lem}

We use the volume argument similar to the one presented in~\cite{eda}.

Let $N=2^{n},$ $\rho$ be the distance induced by Frobenius norm and
$\mu$ be the Haar measure on $\mathbb{U}\left(N\right)$. For the unitary
$U$ we define the volume of its $\varepsilon$-neighbourhood as: 
\[
v\left(U,\varepsilon\right)=\mu\left\{ V \in \mathbb{U}\left(N\right)|\rho\left(M_{V},U\right)\le\varepsilon\right\} .
\]
Let $G^{k}$ be the set of all unitaries that can be constructed
using $k$ gates from the library $G$. Suppose that for any unitary $V$ we can
find a unitary $U$ from $G^{k}$ within the distance $\varepsilon$
from $M_{V}$. This implies: 
\[
\mu\left(\mathbb{U}\left(N\right)\right)\le\sum_{U\in G^{k}}v\left(U,\varepsilon\right)\le\left|G\right|^{k}\max_{U\in G^{k}}v\left(U,\varepsilon\right).
\]
We will show that the volume $v\left(U,\varepsilon\right)$ is upper bounded by $C_{0}\varepsilon^{N^2}$, for some constant $C_0$, therefore: 
\begin{equation}
k\ge\frac{1}{\log\left|G\right|}\log\left(\frac{\mu\left(\mathbb{U}\left(N\right)\right)}{C_{0}\varepsilon^{N^{2}}}\right).\label{eq:nest-1}
\end{equation}
We next show how to estimate $v\left(U,\varepsilon\right)$.  Let $U_{0}$
be a submatrix of $U$ defined as follows: 
\[
U_{0}:=\left\{\bra{e_{i}}\otimes\bra 0\right)U\left(\ket 0\otimes\ket{e_{j}}\right\}
\]
where $\left\{\ket{e_{i}}\right\}$ is the standard (computational) basis in $\mathbb{C}(N)$.
Taking into account that the distance $\rho$ is induced by Frobenius
norm, we write $\rho\left(U,M_{V}\right)\ge\rho\left(U_{0},V\right)$. Therefore: 
\[
v\left(U,\varepsilon\right)=\mu\left\{ V|\rho\left(M_{V},U\right)<\varepsilon\right\} \le\mu\left\{ V|\rho\left(U_{0},V\right)<\varepsilon\right\} .
\]
Let us define $V_{min}$ to be a unitary closest to $U_{0}$.  To
estimate $v\left(U,\varepsilon\right)$ it suffices to consider the case when $\rho\left(V_{min},U_{0}\right)<\varepsilon$. 
The distance $\rho$ is unitarily invariant, therefore $\rho\left(V_{min}^{\dagger}U_{0},I\right)<\varepsilon$
and 
\[
\left\{ V|\rho\left(U_{0},V\right)<\varepsilon\right\} =\left\{ V|\rho\left(V_{min}^{\dagger}U_{0},V\right)<\varepsilon\right\} .
\]
From the triangle inequality,
\[
\rho\left(I,V\right)\le\rho\left(V_{min}^{\dagger}U_{0},I\right)+\rho\left(V_{min}^{\dagger}U_{0},V\right),
\]
we conclude that
\[
\left\{ V|\rho\left(V_{min}^{\dagger}U_{0},V\right)<\varepsilon\right\} \subseteq\left\{ V|\rho\left(I,V\right)<2\varepsilon\right\}.
\]
Finally, 
\[
v\left(U,\varepsilon\right)\le\mu\left\{ V|\rho\left(I,V\right)<2\varepsilon\right\}.
\]
As shown in~\cite{eda}, there exists a constant $C_{0}$ such that the volume
of the ball $\left\{ V|\rho\left(I,V\right)<2\varepsilon\right\}$
is less than $C_{0}\varepsilon^{N^{2}}$.

Estimate (\ref{eq:nest-1}) on $k$ shows that we need circuits of
the size at least $\Omega(\log(1/\varepsilon))$ to cover the full group $\mathbb{U}\left(N\right)$.
If $k$ is chosen in such a way that the inequality (\ref{eq:nest-1})
does not hold, due to the volume argument, there exists a unitary $V\left(\varepsilon\right)$
such that it is not possible to approximate any unitary from $M_{V\left(\varepsilon\right)}$
with precision $\varepsilon$ using at most $k$ gates. 

\section{Future work}

There are some interesting questions that remain to be answered.
The first one concerns the practicality of the proposed construction.  In particular, what are the
constants hidden behind the big-$O$ notation in our approach, and can they be optimized (while further optimizations are only possible up to a multiplicative factor
they are, nevertheless, important for practical purposes)?  The original
algorithm proposed in~\cite{mqes} uses a decomposition into single and
two level unitaries. Each single and two level unitary may have a relatively
large (yet, resulting in a blow up by at most a constant factor,~\cite{mtm}) implementation cost.  An example is given by the CNOT gate, whose controlled version, the Toffoli gate, requires a strictly positive number of T gates, whereas none are needed for constructing the CNOT itself.  
Furthermore, T gate is known to be more difficult to implement fault tolerantly compared to any of the Clifford gates.  
Next, what are the possible trade-offs between adding/reducing ancillae and the gate count? 
Is it possible to use
other efficiently solvable Diophantine equations to discover approximations of other types of gates?  Lastly, does there exist an efficient
algorithm to round off single-qubit unitaries to those single-qubit unitaries over the ring $\mathbb{Z}\left[i,\frac{1}{\sqrt{2}}\right]$
and avoid the need for ancillary qubits altogether? 

\section{Acknowledgments}
Authors supported in part by the Intelligence Advanced Research Projects
Activity (IARPA) via Department of Interior National Business Center
Contract number DllPC20l66. The U.S. Government is authorized to reproduce
and distribute reprints for Governmental purposes notwithstanding
any copyright annotation thereon. Disclaimer: The views and conclusions
contained herein are those of the authors and should not be interpreted
as necessarily representing the official policies or endorsements,
either expressed or implied, of IARPA, DoI/NBC or the U.S. Government.

Michele Mosca is also supported by Canada's NSERC, MPrime, CIFAR,
and CFI. IQC and Perimeter Institute are supported in part by the
Government of Canada and the Province of Ontario.

We wish to thank Alex Bocharov, Martin Roetteler, and Peter Selinger for their comments and helpful discussions.
\nocite{*}


\begin{thebibliography}{10}
\providecommand{\url}[1]{#1}
\csname url@samestyle\endcsname
\providecommand{\newblock}{\relax}
\providecommand{\bibinfo}[2]{#2}
\providecommand{\BIBentrySTDinterwordspacing}{\spaceskip=0pt\relax}
\providecommand{\BIBentryALTinterwordstretchfactor}{4}
\providecommand{\BIBentryALTinterwordspacing}{\spaceskip=\fontdimen2\font plus
\BIBentryALTinterwordstretchfactor\fontdimen3\font minus
  \fontdimen4\font\relax}
\providecommand{\BIBforeignlanguage}[2]{{%
\expandafter\ifx\csname l@#1\endcsname\relax
\typeout{** WARNING: IEEEtran.bst: No hyphenation pattern has been}%
\typeout{** loaded for the language `#1'. Using the pattern for}%
\typeout{** the default language instead.}%
\else
\language=\csname l@#1\endcsname
\fi
#2}}
\providecommand{\BIBdecl}{\relax}
\BIBdecl

\bibitem{Barenco1995}
\BIBentryALTinterwordspacing
A.~Barenco, C.~Bennett, R.~Cleve, D.~DiVincenzo, N.~Margolus, P.~Shor,
  T.~Sleator, J.~Smolin, and H.~Weinfurter, ``{Elementary gates for quantum
  computation},'' \emph{Physical Review A}, vol.~52, no.~5, pp. 3457--3467,
  November 1995. \url{http://arxiv.org/abs/quant-ph/9503016}
\BIBentrySTDinterwordspacing

\bibitem{kbook}
A.~Y. Kitaev, A.~H. Shen, and M.~N. Vyalyi, \emph{{Classical and Quantum
  Computation}}, ser. Graduate studies in mathematics, v. 47.\hskip 1em plus
  0.5em minus 0.4em\relax Boston, MA, USA: American Mathematical Society, 2002.

\bibitem{skalg}
\BIBentryALTinterwordspacing
C.~M. Dawson and M.~A. Nielsen, ``{The Solovay-Kitaev algorithm},''
  \emph{Quantum Information {\&} Computation}, vol.~6, no.~1, pp. 81--95, May
  2005. \url{http://arxiv.org/abs/quant-ph/0505030}
\BIBentrySTDinterwordspacing

\bibitem{eda}
\BIBentryALTinterwordspacing
A.~W. Harrow, B.~Recht, and I.~L. Chuang, ``{Efficient discrete approximations
  of quantum gates},'' \emph{Journal of Mathematical Physics}, vol.~43, no.~9,
  pp. 4445--4451, November 2002.
  \url{http://arxiv.org/abs/quant-ph/0111031}
\BIBentrySTDinterwordspacing

\bibitem{sqlg}
\BIBentryALTinterwordspacing
A.~G. Fowler, ``{Constructing arbitrary Steane code single logical qubit
  fault-tolerant gates},'' \emph{Quantum Information {\&} Computation},
  vol.~11, no.~9, p.~8, November 2011.
  \url{http://arxiv.org/abs/quant-ph/0411206}
\BIBentrySTDinterwordspacing

\bibitem{chem}
\BIBentryALTinterwordspacing
N.~C. Jones, J.~D. Whitfield, P.~L. McMahon, M.-h. Yung, R.~{Van Meter},
  A.~Aspuru-Guzik, and Y.~Yamamoto, ``{Simulating chemistry efficiently on
  fault-tolerant quantum computers},'' April 2012.
\url{http://arxiv.org/abs/1204.0567}
\BIBentrySTDinterwordspacing

\bibitem{Kliuchnikov2012}
\BIBentryALTinterwordspacing
V.~Kliuchnikov, D.~Maslov, and M.~Mosca, ``{Fast and efficient exact synthesis
  of single qubit unitaries generated by Clifford and T gates},'' June 2012.
  \url{http://arxiv.org/abs/1206.5236}
\BIBentrySTDinterwordspacing

\bibitem{mqes}
B.~Giles and P.~Selinger, ``{Exact synthesis of multi-qubit Clifford+T
  circuits},'' December 2012. \url{http://arxiv.org/abs/1212.0506}

\bibitem{mtm}
\BIBentryALTinterwordspacing
M.~Amy, D.~Maslov, M.~Mosca, and M.~Roetteler, ``{A meet-in-the-middle
  algorithm for fast synthesis of depth-optimal quantum circuits},''
  June 2012. 
  \url{http://arxiv.org/abs/1206.0758}
\BIBentrySTDinterwordspacing

\bibitem{Rabin1986}
M.~O. Rabin and J.~O. Shallit, ``{Randomized algorithms in number theory},''
  \emph{Communications on Pure and Applied Mathematics}, vol.~39, no.~S1, pp.
  S239--S256, 1986.

\end{thebibliography}

\end{document}